\documentclass[twocolumn,showpacs,titlepage,aps,prd,letter,
tightenlines,amsmath,byrevtex,nofootinbib,floatfix]{revtex4}
\usepackage{graphicx}
\begin{document}
\title{Detection of Geoneutrinos:\\
Can We Make the Gnus Work for Us?}

\author{John G. Learned}

\address{Department of Physics and Astronomy, University of Hawaii,\\ 
2505 Correa Road, Honolulu, HI 96822 USA}

%\ead{jgl@phys.hawaii.edu}

\begin{abstract} 

The detection of electron anti-neutrinos from natural radioactivity in the earth has been a goal 
of neutrino researchers for about half a 
century\cite{eder_1966,marx_1969,avilez_1981,krauss_1984,raghavan_1998,rothschild_1998}. It was 
accomplished by the KamLAND Collaboration in 2005\cite{araki_2005b}, and opens the way towards 
studies of the Earth's radioactive content, with very important implications for geology.  New 
detectors are operating (KamLAND\cite{decowski_2008} and Borexino\cite{galbiati_2008}), building 
(SNO+\cite{robertson_2008}) and being proposed (Hanohano, LENA, Earth and others) that will go 
beyond the initial observation and allow interesting geophysical and geochemical research, in a 
means not otherwise possible.  Herein we describe the approaches being taken (large liquid 
scintillation instruments), the experimental and technical challenges (optical detectors, 
directionality), and prospects for growth of this field.  There is related spinoff in particle 
physics (neutrino oscillations and hierarchy determination), astrophysics (solar neutrinos, 
supernovae, exotica), and in the practical matter of remote monitoring of nuclear reactors.

\end{abstract}

\maketitle

\section{Introduction: Geoneutrino Studies Started}

The preceding paper, ``Why Geoneutrinos are Interesting"\cite{mcdonough_2008}, really sets the 
stage for this contribution to NU2008.  McDonough explains that the flux of geoneutrinos coming 
mainly from the natural radioactive decay chains of Uranium and Thorium from throughout the earth, 
serves as a tag for the abundance and location of these rare isotopes. While much of this 
radioactive material is surely in the relatively thin continental crusts, roughly one half resides 
in the mantle.  Quite surprisingly to many physicists, these most detectable of the natural decay 
neutrinos are thought to originate in the decay chains that constitute the major source of the 
Earth's internal heating.  That heating in (or under) the mantle of course is responsible for all 
of the mantle circulation which produces continental drift, seafloor spreading, mid-ocean volcanoes 
and traveling hot spots, and of course earthquakes and tsunamis.  Moreover, the geomagnetic field 
is thought to be produced in the outer core of the earth, which is liquid (as we know from the lack 
of seismic shear waves), largely composed of Iron and Nickel, and which convects much faster than 
the viscous mantle material (mostly silicates).  There is not much certain about the depth and 
configuration of the mantle convection, nor of the lateral homogeneity of the U/Th abundances.  
Moreover, there is no consensus about the exact magnetohydrodynamical processes in operation to 
produce the geomagnetic fields, which do indeed change significantly on a human timescale though 
having been present for billions of years.

In principle one can perform a kind of tomography with neutrinos to map out the distribution of the 
sources.  The job is however a tough one, being that thousand ton scale detectors (e.g. KamLAND) 
are required to even begin first detection measurements.  Moreover the high sensitivity required 
demands employment of expensive liquid scintillators in order to produce significant light 
(yielding 30-50 times that from a Cherenkov detector) and further, almost all neutrino 
directionality is lost. Further, delicate care must be taken for radiopurity, now well understood 
but not easy.

The process employed for detection of the anti-neutrinos is the inverse beta decay, used by 
researchers since the initial observations of these neutrinos by Cowan and Reines in the 1950's. 
The signature consists of two flashes of light, near in time and space, and of similar amplitude.  
The first flash is due to the annihilation of the positron which results from a (free) proton being 
struck by an electron-antineutrino (one can think of it as the neutrino stealing a charge from the 
proton).  The neutron is then free to wander about until it captures on another proton to form 
Deuterium, with the liberation of the 2.2 MeV binding energy.  The primary interaction has a 
threshold of the proton-neutron mass difference, and is 1.3 MeV.  Hence the key geonu signature is 
the detection of a primary flash equivalent to a neutrino energy between roughly 1.3 and 3.6 MeV 
(and consisting of a thousand or so photons), a second flash equivalent to 2.2 MeV and delayed by 
about 200 microseconds, and everything originating in a region on the order of one meter in size in 
the detector.  This forms a beautiful discriminant against non-anti-neutrino backgrounds of order 
$10^9$ (depending on size, depth and other factors, including rejection of solar neutrinos which 
make only one flash).

\section{Various Experiments}

The following table indicates the operating, soon to be operating, and proposed experiments of 
relevance around the world. KamLAND has been operating for 6 years now, and is described in the 
talk of Decowski\cite{decowski_2008}.  It was the first to report detection of neutrinos from the 
earth, as was done in a cover article in Nature in 2005\cite{araki_2005b}.  This detection, now 
improved since publication, presented the first demonstration of the expected signal from 
terrestrial radioactivity and, though feeble in statistical power, is roughly in agreement with 
expectations (to order 20\%).

Unfortunately measurements on the continental crust will mostly measure only those neutrinos 
originating in the crust in the detector's neighborhood (500 km or so), with the mantle originating 
neutrinos only contributing on the order of 25\% to the total.  Since even the predictions of the 
local crustal neutrino flux are uncertain at the 20\% level, one cannot discern the mantle 
contribution from a detector location on or near the continental plates. Nonetheless, measurements 
of the geoneutrino flux from continental locations are interesting, since neither the total amount 
of the U \& Th in the Earth is certain, nor is the distribution between crust and mantle.

The Borexino detector in a tunnel of the Gran Sasso Laboratory in the Apennines in Italy has 
started in 2007 and has solar neutrino data already\cite{galbiati_2008}, but at 100 tons mass 
is too small to make much contribution to the geoneutrino business.

The SNO+ detector\cite{robertson_2008}, as a 1000 ton liquid scintillator conversion of the older 
SNO (heavy water) detector in Subury Canada, will make interesting crustal measurements in a 
location above ancient continental plate.

The LENA detector\cite{rubbia_2008} has been talked about for a few years in Europe as a very 
large, mine based, liquid scintillator detector in the 50-100 kiloton class.  It is now part of 
the trio of detectors being studied as a European Megaproject (MEMPHYS, along with a megaton 
water Cherenkov instrument, Laguna, and a 100 kiloton liquid argon device, Glacier).  Various 
locations have been suggested but the favorite for LENA appears to be in a mine in Pyhalsalmi, 
Finland.  The LENA team is centered in Munich, and they have done many excellent studies of the 
physics and technology for this proposed project.  An option to place LENA underwater co-located 
with the NESTOR Project near Pylos, Greece has been discussed, but appears not to be the main 
plan at this time.

The EARTH project has been presented by a Dutch/South African team, with the goal of putting a 
many-armed detector underground on the Island of Curacao\cite{demeijer_2004}.  The notion is to employ 
long, relatively thin, layered neutrino detectors, getting some directionality from the relative 
rates in each arm.  A detailed detector design has not been presented yet, and performance is as 
yet not well determined.

\begin{center}
\begin{table}[h]
\caption{ List of various geoneutrino experiments.}
%\footnotesize\rm
\centering
\begin{tabular}{@{}*{7}{l}}
%\br
Detector   & Region   & Location & Size     & Status (Start)       \\
%\mr
\ KamLAND  & Japan    & Mine     & 1000 T   & Operating (2002)     \\
\ Borexino & Italy    & Tunnel   &  100 T   & Operating (2007)     \\
\ SNO+     & Canada   & Mine     & 1000 T   & Construction (2010)  \\
\ Hanohano & Pacific  & Ocean    & 10,000 T & Proposed (2013?)     \\
\ LENA     & Finland? & Mine     & 50,000 T & Proposed (?)         \\
\ EARTH    & Curacao  & Drill Holes & ??    & Discussed (??)       \\
%\br
\end{tabular}
\end{table}
\end{center}

\section{Hanohano}

A deep ocean antineutrino observatory called Hanohano (Hawaii Anti-Neutrino Observatory) is being 
developed at Hawaii and with collaborators elsewhere\cite{hanohano}. The observatory will record 
interactions of electron antineutrinos of $E_{\nu} > 1.8 MeV$ by inverse $\beta$-decay in a 
monolithic cylindrical detector of 10 kt of ultra-pure scintillating liquid. An outer surface array 
of inward-looking 10-inch photomultiplier tubes in 13-inch glass pressure housings will collect 
scintillation photons. The planned energy resolution is $3.5\%/\sqrt{E_{vis}}$, with $E_{vis} = 
E_{\nu} - 0.8 MeV$ the visible energy. Sufficient overburden (3 km or more for most sensitive geonu 
studies), adequate shielding (from ocean radioactivity), and radio-pure detector components will 
limit background to negligible levels providing very high detection efficiency.

Considerable science potential derives from the ability to deploy the observatory at various deep 
ocean locations. An initial deployment offshore a nuclear reactor complex for measuring neutrino 
mixing parameters could be followed, for example, by a deployment near Hawaii for measuring 
terrestrial antineutrinos. This flexibility presents a significant advantage over similar 
observatories at a fixed underground locations.

The preliminary design, resulting from a two year engineering study by Makai Ocean Engineering, 
specifies the detector to be a right cylindrical shape, transported in a special barge from which 
it may be deployed, and recovered.  The inner volume of liquid scintillator is separated from the 
photodetectors by a segmented acrylic layer.  The individual optical units are in clusters, in 
plain oil. Outside this stainless steel vessel will be a further layer of 2 meters of pure water, 
and veto photomultipliers (as well this layer provides access to the inner tank for installation 
and repairs).  Very importantly in this design, all fiber-optic and electrical connections will 
be made pierside, tested and calibrated prior to deployment (avoiding previous bad experience 
with unreliable connectors in the ocean, and difficult remote connections via robot).  The 
detector design is aimed at multiple deployments on a roughly annual cycle, allowing changes of 
venue to follow the science.

\subsection{Hanohano Geoneutrino Studies}

The Hanohano team aims for measuring the flux of U/Th geo-neutrinos from earth's mantle with 25\% 
uncertainty in one year of operation near Hawaii\cite{dye_2006a}. Included in this 
statistic-dominated result is 9\% systematic error due to uncertainty in the U/Th content of the 
crusts. This same uncertainty limits the precision of measurements of the mantle flux at 
continental locations to $>50\%$. Not included in the analysis is uncertainty of the neutrino 
mixing angles $\theta_{12}$[3] and $\theta_{13}$\cite{apollonio_2003}. The survival probability 
for fully mixed geo-neutrinos is 59\% (+6\%, - 15\%), as measured at present. The upper (lower) 
value obtains with minimum (maximum) present values of the mixing angles. Imprecise knowledge of 
mixing angles and U/Th content of earth's crusts introduce comparable uncertainties to the 
measurement by Hanohano of geo-neutrinos from the mantle. Nonetheless, deployments at several 
widely-spaced mid-ocean locations test lateral heterogeneity of uranium and thorium in the 
mantle.

Geo-neutrinos with energy between 1.8 MeV and 2.3 MeV come from both $^{238}U$ and $^{232}Th$ 
decay products, while those between 2.3 MeV and the maximum energy of 3.3 MeV are only from the 
$^{238}U$ decay product $^{214}Bi$. This spectral feature allows a measurement of the Th/U ratio. 
Although geology traditionally ascribes the chondritic Th/U ratio of $3.9\pm 0.1$ to the bulk 
earth, samples from the upper mantle reveal a substantially lower value of 2.6 suggesting layered 
mantle convection\cite{turcotte_2001}. Geo-neutrino flux measurements sample large volumes of the 
deep earth providing an important test of mantle convection models.

An earth-centered natural fission reactor\cite{herndon_1996,hollenbeck_2001} is a speculative, 
untested hypothesis. Predicted to be in the power range of 1-10 TW, it has the potential to 
explain the variability of the geo-magnetic field and the anomalously high helium-3 
concentrations in hot-spot lavas. Fission products from such a geo-reactor would undergo 
$\beta$-decay, producing antineutrinos with the characteristic nuclear reactor spectrum. A 
one-year deployment of Hanohano at a mid-ocean location well distant from nuclear power plants 
tests the existence of the geo-reactor. This deployment would set a 99\% CL upper limit to the 
geo-reactor power at 0.3 TW or, were a 1 TW geo-reactor to exist, produce nearly a $5\sigma$ 
measurement\cite{dye_2006b}.

\subsection{Hanohano Neutrino Studies}

Neutrino mixing and oscillation\cite{mohapatra_2007} are responsible for the deficit of solar 
neutrinos\cite{aharmin_2005}, the spectral distortion of reactor 
antineutrinos\cite{araki_2005a}, and the deficit of atmospheric muon neutrinos\cite{ashie_2005}, 
which has been confirmed using an accelerator-produced muon neutrino beam\cite{aliu_2005}. These 
initial observations reduce the allowed regions of neutrino mixing parameter space, guiding 
future precision measurements of mixing angles and mass-squared differences, including resolution 
of the spectrum of neutrino masses. Positioning Hanohano ~60 km distant from a nuclear reactor 
complex enables precision measurement of $\theta_{12}$ and, for non-zero $\theta_{13}$, $\delta 
{m^2}_{31}$. This latter measurement can lead to a determination of neutrino mass hierarchy.

Several authors discuss a precision measurement of the solar mixing angle $\theta_{12}$ using 
antineutrinos from a nuclear reactor\cite{bandyopadhyay_2003,bandyopadhyay_2005,minakata_2005}. 
The experiment utilizes a near detector at the reactor complex for normalizing flux and cross 
section with a far detector at the first minimum of survival probability. There is agreement that 
an exposure of 60 GW-kt-y of a far detector at a distance of ~60 km yields an uncertainty of 2\% 
in the value of $sin^2(\theta_{12})$ at the 68\% confidence level, assuming a detector systematic 
uncertainty of 4\% or less. This experiment is analogous to methods proposed for precision 
measurement of the sub-dominant mixing angle $\theta_{13}$\cite{anderson_2004}, namely the 
reactor antineutrino flux sampling defines one-half cycle of the oscillating survival 
probability. The difference is the distance of the first minimum of survival probability, which 
is ~2 km for the $\theta_{13}$ measurement.

The far detector for the $\theta_{12}$ experiment records multiple cycles of $\delta {m^2}_{13}$ 
oscillation given that $\theta_{13} > 0$, the detector has adequate energy resolution, and 
sufficient exposure. There is a plan to measure these cycles in L/E space by sampling the Fourier 
power at different values of $\delta m^2$\cite{learned_2008}. This self-normalizing, robust 
method offers a precision measurement of $\delta {m^2}_{13}$ for $sin^2(2\theta_{13}) > 0.05$, 
determines neutrino mass hierarchy by evaluating asymmetry of the Fourier power spectrum, and 
measures $\theta_{13}$; all without the need for a near detector.

Hanohano is capable of performing the experiments described above with a one-year deployment 
offshore a suitable nuclear reactor complex. At least two candidate sites with considerable 
overburden exist. One is in 1100 m of water West of the ~7 GW San Onofre reactors in California 
and the other is in 2800 m of water east of the ~6 GW Maanshan reactors in Taiwan.  Other locations 
are possible.

\subsection{Hanohano Other Physics Opportunities}

The existence of a detector with the capabilities of Hanohano and the location deep inside the 
ocean offers an opportunity for several exciting discoveries. Here we indicate some of them 
briefly. One is a search for an anti-neutrino signal from relic supernovae from the distant past. 
The signal can be as large as 4 events per year\cite{strigari_2004}.  Of course, the signal from 
a galactic supernova is much more robust: about 2000 conventional anti-electron neutrino capture 
events, with additional charged current events from $^{12}N$, $^{12}B$, and neutral current 
events from scattering on electrons and $^{12}C$. There are also about 2000 events of elastic 
scattering on protons. This would be a signal in addition and complimentary to those from all 
other detectors around the world.

If the purity levels in the scintillating liquid can be made as low as in Borexino, then one may 
search for the neutrino signal from solar neutrinos especially the pep line and the CNO 
neutrinos. The signal from pep+CNO in the window of 0.8 to 1.3 MeV is expected to be about 150 
events/day/10kt. This is detectable against an expected background of about 130 events/day/10kt, 
mostly from 11C. This is very useful in confirming the conventionally accepted neutrino 
parameters and to rule out some non-standard neutrino properties\cite{friedland_2004}.

Finally, Hanohano has the capability to detect the proton decay mode $p \rightarrow \nu + K^+$. 
This is expected to occur in super-symmetric models. Liquid scintillation detectors an advantage 
over conventional water-Cherenkov detectors by being able to detect the kaon directly. Hanohano 
can reach a sensitivity of almost $10^{34}$ y.

\section{The Sad Story of K40 Detection}

Geologists would very much like to detect the neutrinos from Potassium-40 
decay\cite{mcdonough_2008}.  This common isotope has different chemical properties than the heavy 
elements, and is thus to be found in different regions of the Earth, from the oceans and crust to 
the core.  The unfortunate case is that the end point energy (1.3 MeV) is below the inverse $\beta$ 
threshold, and hence the signature of these neutrinos' interactions is only one flash of light, to 
be extracted from the solar neutrino and other backgrounds.  We will not say more here, other than 
to report it as a major challenge for which good ideas are needed\cite{krauss_1984}.

\section{Towards Directionality}

A great challenge for a next generation of low energy anti-neutrino detectors is achieving some 
directional resolution.  People have discussed tracking detectors, but the sizes required are 
formidable and not yet practical. The nearest opportunity seems to be with the inverse-$\beta$ 
decay itself.  The neutron acquires a tiny (few keV/c) amount of momentum from the striking 
neutrino.  If one can record the positron appearance location and the neutron absorption location, 
one can make a poor ($O(20^o)$) angular determination.  Improvements in present technology can come 
through better vertex resolution, track resolution, shorter scintillator emission times, heavier 
materials (shorter gamma travel distances), and greater neutron cross sections.  Employment of an 
alpha emitter can help in locating the neutron absorption, but the alpha's suffer from greatly 
decreased light yield due to saturation of the scintillator by heavily ionizing particles. Imaging 
can help by permitting one to recognize the annihilation gamma's topology and getting a better 
initial neutrino interaction vertex.  Ditto for the neutron absorption location.  Advances in CCDs 
and optics may make some significant progress here, but not immediately.  Studies are underway at 
several institutions (particularly at RCNS, Tohoku) to push ahead in this area.

\section{Other Applications}

Due to lack of space we cannot elaborate upon the practical applications of these detectors.  An 
introduction is given in Bowden\cite{bowden_2008}, where the focus is upon close-in reactor 
monitoring.  Remote monitoring of reactor activity will be possible out to hundreds of 
kilometers with next generation instruments, and some can envisage a worldwide 
network of neutrino monitors contributing to anti-nuclear weapons proliferation efforts in the 
future\cite{dye_2007}.

\section{Summary}

The business of geoneutrino detection is already underway at existing large instruments and those 
being built.  The first results are in hand, and they will continue to improve.  The most 
desirable information about U and Th content of the mantle (and core) will not be obtained until a 
10 kiloton scale deep ocean instrument is deployed.  Hanohano is proposed for this task, but is 
at least several years from operations.  Studies are underway for further improvements in 
detector sensitivity and directional capability.

Operation of the class of multi-kiloton neutrino detectors with MeV level sensitivities will not 
only open a new area in geology, but will make contributions to neutrino physics and astrophysics, 
and will pave the path towards the creation of networks of detectors for remote nuclear reactor 
monitoring.  It would appear that there is a bright future in this emerging field.

\subsection{Acknowledgments} 

I want to acknowledge help from many others in preparing this survey.  Particular thanks to Steve 
Dye, Sanshiro Enomoto, Gene Guillian, Eligio Lisi, Bill McDonough, Sandip Pakvasa, Bob Svoboda, 
Nikolai Tolich, and members of the DUSEL, Hanohano, KamLAND, LENA, and SNO groups for graphics and 
physics help. This work was partially funded by U.S. Department of Energy grant DE-FG02-04ER41291 
and the University of Hawaii. We would like to acknowledge our collaborators on the Hanohano 
Collaboration Council:  A. Bernstein, M. Cribier, F. von Feilitzsch, G. Horton-Smith, K. Inoue, T. 
Lasserre, J. Mahoney, J. Maricic, W. McDonough, S. Parke, A. Piepke, R. Raghavan, N. Solomey, R. 
Svoboda and J. Wilkes.

\end{document}